%% file: inducedphin.tex
\documentclass[a4paper,11pt]{article}
\pdfoutput=1 % if your are submitting a pdflatex (i.e. if you have
\usepackage{jcappub} % for details on the use of the package, please
\usepackage{hyperref}

\usepackage[T1]{fontenc} % if needed
\usepackage{footnote}
\def\be{\begin{equation}}
\def\ee{\end{equation}}
\def\ba{\begin{eqnarray}}
\def\ea{\end{eqnarray}}

\usepackage{array}

\title{\boldmath{Cosmological constraints on induced gravity dark energy models}}

\author[a,b,c]{M. Ballardini}
\author[b,c]{F. Finelli}
\author[d,e,f]{C. Umilt\`a}
\author[b,c]{and D. Paoletti}

\affiliation[a]{DIFA, Dipartimento di Fisica e Astronomia, Alma Mater Studiorum Universit\`a 
di Bologna, Viale Berti Pichat 6/2, I-40127 Bologna, Italy}
\affiliation[b]{INAF-IASF Bologna, via Gobetti 101, I-40129 Bologna, Italy}
\affiliation[c]{INFN, Sezione di Bologna, Via Irnerio 46, I-40126 Bologna, Italy}
\affiliation[d]{Institut d'Astrophysique de Paris, CNRS (UMR7095), 98 bis Boulevard 
Arago, F-75014, Paris, France}
\affiliation[e]{UPMC Univ Paris 06, UMR7095, 98 bis Boulevard Arago, F-75014, Paris, France}
\affiliation[f]{Sorbonne Universit\'es, Institut Lagrange de Paris (ILP), 98 bis Boulevard 
Arago, 75014 Paris, France}

\emailAdd{ballardini@iasfbo.inaf.it}
\emailAdd{finelli@iasfbo.inaf.it}
\emailAdd{umilta@iap.fr}
\emailAdd{paoletti@iasfbo.inaf.it}

\abstract{We study induced gravity
dark energy models coupled with a simple monomial potential $\propto \sigma^n$ and a positive 
exponent $n$. These simple potentials lead to viable dark energy models with a weak dependence 
on the exponent, which characterizes the accelerated expansion of the cosmological model in 
the asymptotic attractor, when ordinary matter becomes negligible.
We use recent cosmological data to constrain the coupling $\gamma$ to the Ricci curvature, 
under the assumptions that the scalar field starts at rest deep in the radiation era and that 
the gravitational constant in the Einstein equations is compatible with the one measured in 
a Cavendish-like experiment.
By using $Planck$ 2015 data only, we obtain the 95~\% CL bound $\gamma < 0.0017$ for $n=4$, 
which is further tightened to $\gamma < 0.00075$ by adding Baryonic Acoustic Oscillations 
(BAO) data. This latter bound improves by $\sim 30\,$\% the limit obtained with the $Planck$ 
2013 data and the same compilation of BAO data. 
We discuss the dependence of the $\gamma$ and $\dot G_N/G_N (z=0)$ on $n$.}

\begin{document}
\maketitle
\flushbottom

\section{Introduction}
\label{sec:intro}

The most recent data from {\sc Planck} \cite{Adam:2015rua,Ade:2015xua} are consistent with a 
cosmological constant and cold dark matter (CDM) as the dark components which contribute to 
the 96~\% of the energy budget of our Universe. Several dark energy/modified gravity models 
alternative to $\Lambda$CDM model have been compared extensively to the most recent $Planck$ 
data \cite{Ade:2015rim}. 
The predictions for CMB anisotropies and for other cosmological observabels for these modified 
gravity scenarios can be studied either in a model-dependent way or by more general 
{\em parametrized deviations} from Einstein gravity in the metric perturbations \cite{Zhao:2011te}. 
By using both approaches, no compelling evidence in favour of alternative models to $\Lambda$CDM 
have been found in \cite{Ade:2015rim}, in particular when $Planck$ lensing is included 
\cite{Plancklensing15}. Other parametrized deviations from Einstein gravity in extended 
cosmological models have been studied in \citep{DiValentino:2015bja}.

In this paper we study the simplest scalar-tensor dark energy models based on induced gravity (IG) 
\cite{zee} - or Brans-Dicke-like  models \cite{BD} by a redefinition of the scalar field - with 
a monomial potential, extending previous works based on a quartic potential 
\cite{CV,Wetterich,FTV,Umilta:2015cta} (see also 
\cite{Uzan:1999ch,Perrotta:1999am,Bartolo:1999sq,Amendola:1999qq,Chiba:1999wt,Boisseau:2000pr} 
for important works on scalar-tensor dark energy). 
By assuming a monomial potential with a positive exponent $n$ - i.e. $V(\sigma) \propto \sigma^n$ 
- mimicking an effective cosmological constant at recent times, the scalar field is frozen 
during the radiation era and is driven by the non-relativistic components after the 
matter-radiation equality to higher values: in these models the effective Planck mass therefore 
increase in time. 
Such monomial potentials with a positive power are easily motivated at fundamental level 
or within particle physics, as happens for the analogous case of a non-minimally coupled scalar 
field \cite{Chiba:2010cy}.

We extend our self-consistent approach in which we solve simultaneously the background and the 
linear perturbation dynamics in IG previously applied to a quartic potential \cite{Umilta:2015cta}. 
This approach is complementary to the study of parametrized deviations from Einstein gravity 
in metric perturbations, since it allows to accurately study modified gravity theoretical 
models which are arbitrary close to the flat $\Lambda$CDM, without any approximation in the 
background or in the perturbations. We then use recent cosmological data to constrain these 
IG models with a monomial potential. In a previous study we obtained a 95~\% CL bound 
$\gamma < 0.0012$ for a quartic potential by the $Planck$ nominal mission temperature data 
in combination with a BAO compilation \citep{Umilta:2015cta}.

Our paper is organized as follows. In the next section we discuss the background dynamics and 
the dependence of the self-accelerating solutions on the exponent of the monomial potential. 
In section~\ref{sec:fluctuations} we discuss the evolution of linear fluctuations and we test 
our numerical treatment against the analytic approximations in the quasi-static regime, as done 
for the quartic potential in \cite{Umilta:2015cta}. 
We show the dependence of the CMB anisotropies power spectra in temperature and polarization 
on the power of the monomial potential in section~\ref{sec:predictions}. We then use the 
$Planck$ 2015 \cite{Plancklike15,Plancklensing15} and BAO 
\cite{Beutler:2011hx,Ross:2014qpa,Anderson:2013zyy} data to constrain these models in 
section~\ref{sec:Planck2015} and \ref{sec:Planck2015pot}.
We draw our conclusions in section~\ref{sec:concl}.

\section{Dark Energy with a monomial potential within Induced Gravity}
\label{sec:background}

The model we consider is described by the following Lagrangian:
\begin{equation}
S = \int d^4x \sqrt{-g}\, \Bigl[ \frac{\gamma \sigma^2 R}{2} - \frac{g^{\mu \nu}}{2}
\partial_{\mu} \sigma \partial_{\nu} \sigma - V(\sigma) + \mathcal{L}_m \Bigr] 
\label{eqn:IGaction}
\end{equation}
with
\begin{equation}
V (\sigma) = \lambda_n \sigma^n \,.
\label{monomial}
\end{equation}

The Friedmann and the Klein-Gordon equations for IG in a flat Robertson-Walker metric
are respectively:
\begin{equation}\label{fried-ig}
H^2 + 2 H \frac{\dot \sigma}{\sigma} = \frac{\sum_i \rho_i +V(\sigma)}{3\gamma \sigma^2} 
+ \frac{{\dot \sigma}^2}{6 \gamma \sigma^2} 
\end{equation}
\begin{equation}\label{kg-ig}
\ddot{\sigma} + 3H{\dot \sigma}+\frac{{\dot \sigma}^2}{\sigma} 
+ \frac{1}{(1+6\gamma)} \Bigl( V_{,\sigma}-\frac{4V}{\sigma}\Bigr) 
= \frac{1}{(1+6\gamma)} \frac{\sum_i(\rho_i -3p_i)}{\sigma}
\end{equation}
once the Einstein trace equation: 
\begin{equation}
- \gamma \sigma^2 R = T - (1+6 \gamma) \partial_\mu \sigma \partial^\mu \sigma 
- 4 V - 6 \gamma \sigma \Box \sigma
\end{equation}
is used. In the above $V_{,\sigma}$ denotes the derivative of the potential $V(\sigma)$ with respect 
to $\sigma$, the index $i$ runs over all fluid components, i.e. baryons, cold dark matter (CDM), 
photons and neutrinos, and we use a dot for the derivative with respect to the cosmic time. 
The effective potential in Eq.~\ref{kg-ig} vanishes for $n=4$.

In absence of matter, exact solutions with an accelerated expansion exist for this class of monomial 
potentials in Eq.~\ref{monomial} within induced gravity \cite{CFTV} (for earlier works see 
Ref.~\cite{BM}). Solutions with $a (t) \sim t^p$ (with $t>0$ and $p>1$) exist:
\ba
p = 2\frac{1+ (n+2) \gamma }{(n-4)(n-2) \gamma} \,,\,\,
\, \, \sigma(t) = \frac{c_0}{t^\frac{2}{(n-2)}}
\label{scaling_sigma} \,,
\ea
with $4<n<4+\sqrt{2(6+1/\gamma)}$ or $4-\sqrt{2(6+1/\gamma)}<n<2$ and $c_0$ is an integration constant. 
The special cases with $n=2 \,, 4$ (which correspond to poles in the above equations) 
correspond to a de Sitter solution having $a(t) \propto e^{H t}$. However for these two special cases, 
the time evolution for the scalar field are different, being $\sigma$ time dependent for $n=2$ and 
constant in time for $n=4$.

As in \cite{Umilta:2015cta}, we consider the case in which the scalar field $\sigma$ at rest deep 
in the radiation era, since an initial non-vanishing time derivative would be rapidly dissipated 
\cite{FTV}. For values for the parameters of the potential leading to viable models of dark energy, 
the scalar is effectively massless during the radiation and most of the matter era. The scalar field 
$\sigma$ starts evolving form its initial value $\sigma_i$ due the presence of non-relativistic matter:
\be
\sigma (\tau) = \sigma_i \left( 1 + \frac{3}{2} \gamma \omega \tau + {\cal O} (\tau^2) \right) \,, 
\label{sigmatau}
\ee
where the parameter $\omega$ depends on the relativistic and non-relativistic energy density at present:
\be
\omega = \frac{\rho_{m \, 0}}{\sqrt{3 \gamma \rho_{r \, 0}} (1+6 \gamma) \sigma_i}\,.
\ee
At the same next-to-leading order in $\omega \tau$ for a Universe filled by radiation and matter, 
the scale factor is:
\be
a (\tau) = \frac{\rho_{r \, 0}}{\sqrt{3 \gamma} \sigma_i} \tau 
\left( 1 + \frac{\omega}{4} \tau - \frac{5 \gamma}{16}  \omega^2 \tau^2 + {\cal O} (\tau^3) \right) \,.
\label{atau}
\ee
IG therefore induces at next-to-leading order a correction to the evolution of the scale factor 
similar to the case of a negative curvature \cite{cambnotes}.

As in \cite{Umilta:2015cta}, we consider the present value for the scalar field value consistent 
with the Solar System constraints:
\begin{equation}
\gamma \sigma_0^2 = \frac{1}{8\pi G} \frac{1+8\gamma}{1+6\gamma} \,,
\label{sigma0}
\end{equation}
where $G = 6.67 \times 10^{-8}$ cm$^3$ g$^{-1}$ s$^{-2}$ is the gravitational constant measured 
in laboratory Cavendish-type.

\begin{figure}[t!!]
\centering
\epsfig{file=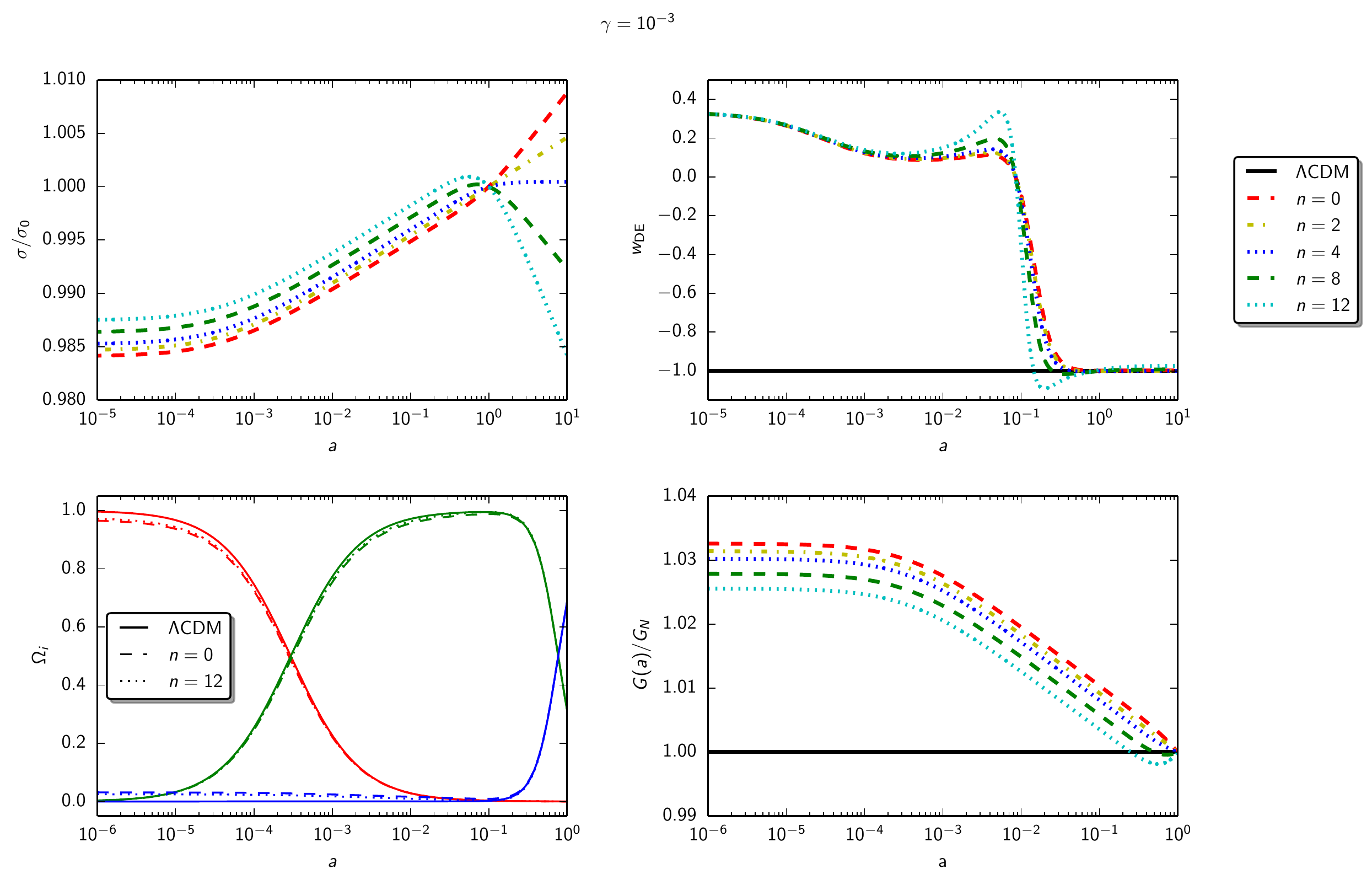, width=15 cm}
\caption{Evolution of $\sigma/\sigma_0$ (upper left panel), $w_{\rm DE}$ (upper right panel) 
and $G_\textup{N} (a)/G \equiv \sigma_0^2/\sigma^2$ (lower right panel)
as function of the scale factor $a$ for $\gamma = 10^{-3}$ and different values of $n$. 
In the lower left panel we show the critical densities $\Omega_i$: radiation in red, matter in green 
and effective dark energy in blue. See text for more details.}
\label{sigmaandw}
\end{figure}

In figure~\ref{sigmaandw} several quantities are plotted versus the scale factor $a$ for 
$\gamma = 10^{-3}$ and for different values of $n$, within the assumption of Eq.~\ref{sigma0}. 
In the upper left panel, the evolution of $\sigma/\sigma_0$ is plotted up to $a=10$ in order to 
show the dependence on $n$ of the future single field attractor of Eq.~\ref{scaling_sigma}. 
In the upper right panel we show the parameter of state $w_\mathrm{DE}$ of the effective dark 
energy component corresponding to Einstein gravity with a Newton's constant given by the 
current value of the scalar field, i.e. $8 \pi G_\textup{N} = (\gamma \sigma_0^2)^{-1}$ as 
introduced in \cite{Boisseau:2000pr}: also in this case we extend the plot up to $a=10$ to show 
that $w_\mathrm{DE} \simeq -1$ for values of $\gamma$ compatible with observations 
\cite{Umilta:2015cta}.
In the lower left panel we show the evolution of the critical densities $\Omega_i$, always 
corresponding to an Einstein gravity system with a Newton's constant given by the current 
value of the scalar field defined in \cite{Umilta:2015cta}. In the lower right panel, we show 
explicitly the evolution of $G_\textup{N} (a)/G \equiv \sigma_0^2/\sigma^2$; it is clear that in 
this class of models the effective Newton's constant (the effective Planck mass 
$M_\textup{pl}^2 (a) = \gamma \sigma^2 (a)$) decreases (increases) with time.

\section{Linear fluctuations and predictions for cosmological observables}
\label{sec:fluctuations}

As in our previous paper for the]]++ quartic potential \cite{Umilta:2015cta}, we consider scalar 
fluctuations in the longitudinal gauge. We refer to \cite{Umilta:2015cta} for the Einstein 
equations in the longitudinal gauge with the substitution of the quartic potential with a generic 
monomial one where necessary.
The equation for the field fluctuations in the longitudinal gauge for a generic potential is:
\begin{equation}
 \begin{split}
  \ddot{\delta \sigma} + \dot{\delta \sigma} \Bigl( 3H +2 \frac{\dot{\sigma}}{\sigma} \Bigr) +
\Bigl[\frac{k^2}{a^2} + \left( V_{\sigma \sigma} + 4 \frac{V}{\sigma^2} - 4 \frac{V_\sigma}{\sigma} \right) 
- \frac{\dot{\sigma}^2}{\sigma^2}+\frac{\sum_i(\rho_i - 3p_i)}{(1+6\gamma)\sigma^2} \Bigr]
\delta \sigma  \\
 =\frac{2\Psi \sum_i(\rho_i - 3p_i)}{(1+6\gamma)\sigma} +\frac{\sum_i(\delta \rho_i -3\delta p_i)}{(1+6\gamma)\sigma}
+\dot{\sigma} \Bigl(3 \dot \Phi + \dot \Psi \Bigr) \,.
 \end{split}
\end{equation}
We note that the terms in the potential and its derivatives in the effective mass of 
$\delta \sigma$ vanish only for $n=4$. We therefore expect a non-trivial dependence of 
$\delta \sigma$ on large scales since the onset of accelerated expansion.

\begin{figure}[t!!]
\centering
\epsfig{file=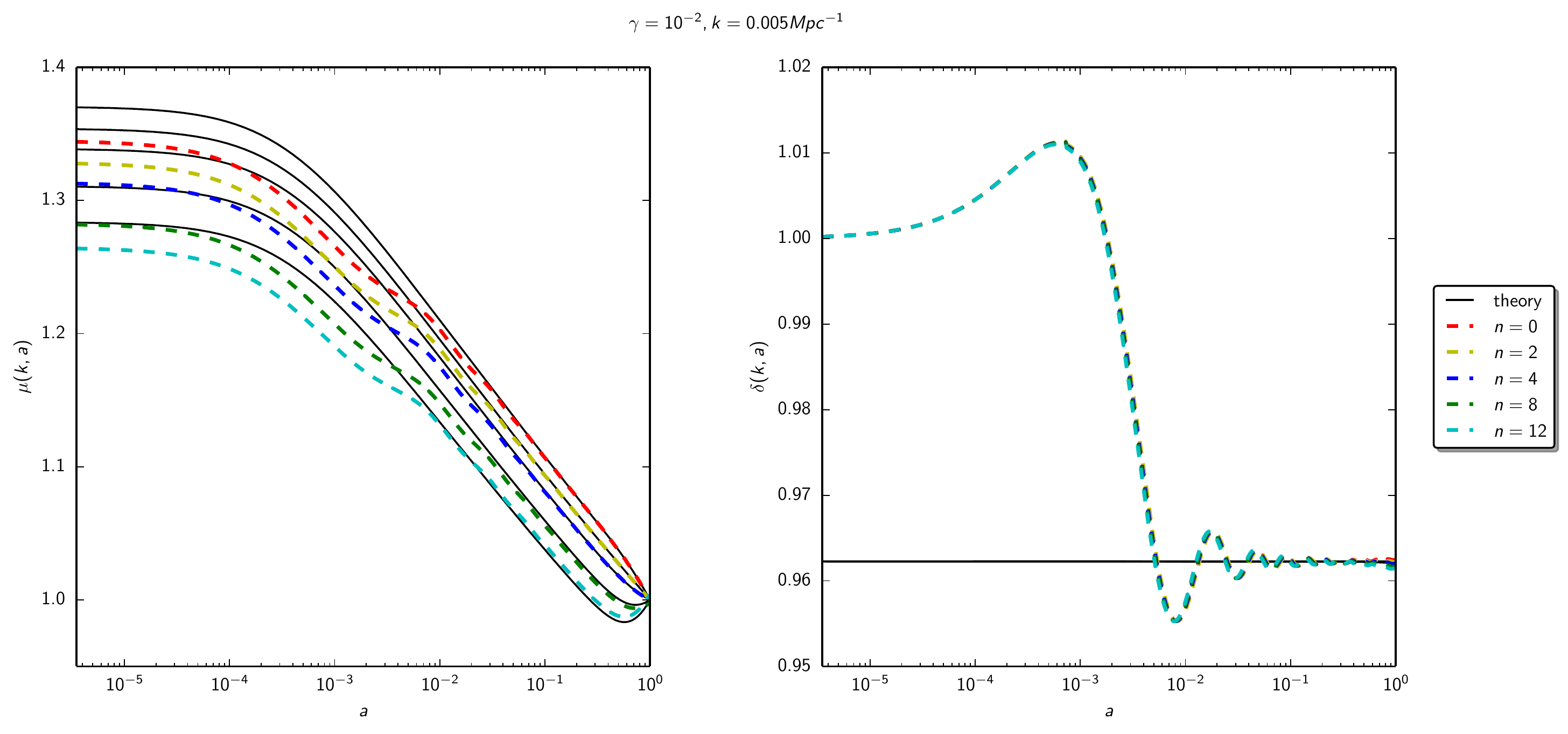, width=15 cm}
\caption{Comparison of the theoretical quasi-static approximations for $\mu (k,a)$ and $\delta(k,a)$ parameters 
(black lines) with our exact numerical results for $k = 0.005\ \mathrm{Mpc}^{-1}$ 
and $\gamma=10^{-2}$ when $n$ is varied.}
\label{mu}
\end{figure}

We have extended our previous modification \cite{Umilta:2015cta} of the publicly available 
Einstein-Boltzmann code CLASS \footnote{\href{www.class-code.net}{www.class-code.net}} 
\cite{Lesgourgues:2011re,Blas:2011rf} to a generic potential.
We therefore test our numerical results obtained by initializing fluctuations in the adiabatic 
initial conditions deep in the radiation era to the quasi-static approximation beyond the case 
of a quartic potential, which was studied previously \cite{Umilta:2015cta}. We consider the 
parameters $\mu (k,a)$ and $\delta (k,a)$:
\begin{eqnarray}
k^2 \Psi &=& - 4 \pi G a^2 \mu (k,a) \left[ \Delta + 3 (\rho+p) \bar{\sigma} \right] \,,
\label{mu2} \\
k^2 [\Phi - \delta(k,a)\Psi] &=&  12 \pi G a^2 \mu(k,a) (\rho+p) \bar{\sigma} \,,
\label{gamma}
\end{eqnarray}
where $\Phi \,, \Psi$ are the Newtonian potentials in the longitudinal gauge, $\Delta$ is the total 
comoving energy perturbation (excluding the contribution from $\sigma$) and $\bar{\sigma}$ is 
the anisotropic stress \cite{Umilta:2015cta}.
In the quasi-static approximation, for generic $n$, the two parameters are approximated as:
\begin{eqnarray}
\mu (k,a) &=& \frac{\sigma_0^2}{\sigma^2} \frac{1+6 \gamma}{1+8 \gamma} 
\frac{1 + 8 \gamma - 2 m^2_\mathrm{eff}/k^2}{1 + 6 \gamma - 2 m^2_\mathrm{eff}/k^2} \nonumber \\
%k^2 \Psi &=& - 4 \pi G a^2 \mu (k,a) \left[ \Delta + 3 (\rho+p) \bar{\sigma} \right] \,,
\label{mu2} \\
\delta (k,a) &=& \frac{1 + 4 \gamma - 2 m^2_\mathrm{eff}/k^2}{1 + 8 \gamma - 2 m^2_\mathrm{eff}/k^2} \nonumber \\ 
%k^2 [\Phi - \delta(k,a)\Psi] &=&  12 \pi G a^2 \mu(k,a) (\rho+p) \bar{\sigma} \,,
\label{gamma}
\end{eqnarray}
where 
\be
m^2_\mathrm{eff} = \frac{d}{d \sigma} \left( \sigma^4 \frac{d}{d \sigma} \left( \frac{V}{\sigma^4} \right) \right) \,.
\ee 
Our exact numerical results are compared with the quasi-static approximation 
in figure~\ref{mu} for $k = 0.005\ \mathrm{Mpc}^{-1}$ and $\gamma=10^{-2}$ when $n$ is varied. 
As already established for the quartic potential \cite{Umilta:2015cta}, the quasi-static 
approximation for $\mu (k, a)$ is quite accurate only for sub-Hubble scales also in the general case, 
i.e. $n \ne 4$. The parameter $\delta (k,a)$ depends on time when $n \ne 4$, but depends on $n$ 
weakly compared to $\mu (k,a)$. 

\section{Predictions for CMB anisotropies and Matter Power spectrum}
\label{sec:predictions}
\begin{figure}[t!!]
\centering
\epsfig{file=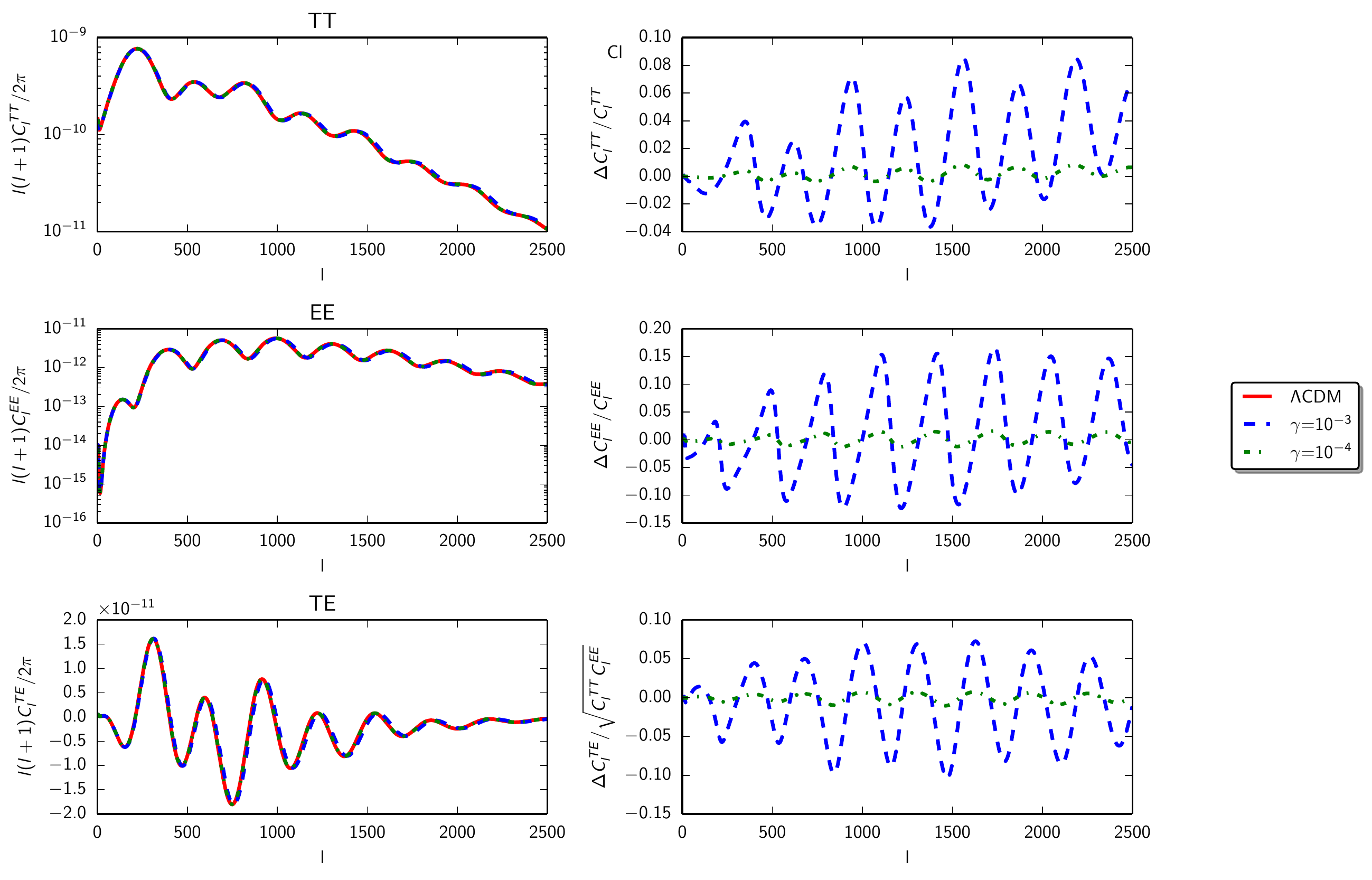, width=15 cm} 
\caption{To the left, from the upper to the lower panel respectively, CMB TT, EE and TE power 
spectra for $\gamma = 10^{-3} \,, 10^{-4}$ and $n=4$. In the upper and middle right panels, we show 
the relative differences for TT and EE spectra with respect to a reference $\Lambda$CDM model. 
In the lower right panel we show the differences for $C_\ell^{TE}$ normalized to 
$\sqrt{C_\ell^{TT} C_\ell^{EE}}$.}\label{fig:clT_delta}
\end{figure}

\begin{figure}[t!!]
\centering
\epsfig{file=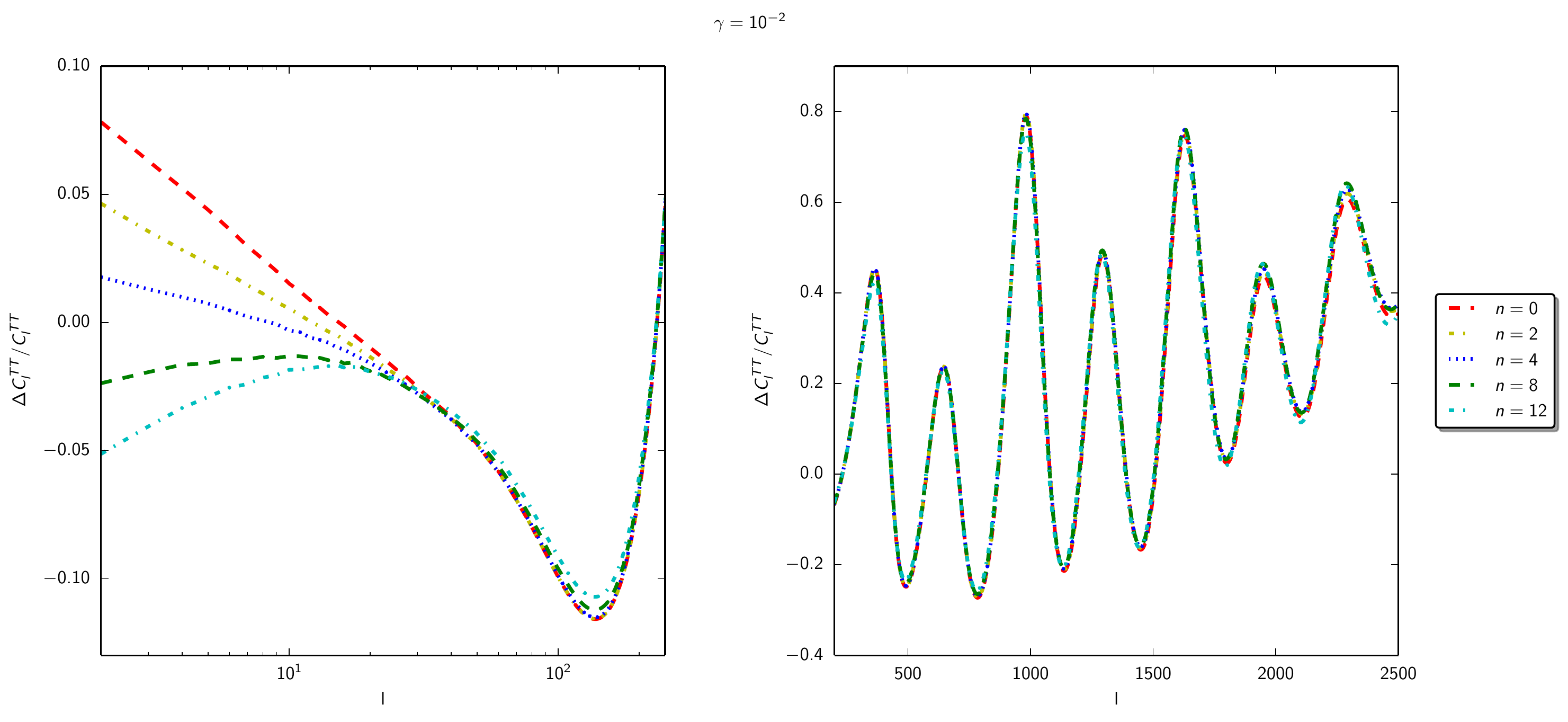, width=15 cm} \\
\caption{Relative differences for the CMB temperature anisotropies power spectrum  
with respect to a reference $\Lambda$CDM for $\gamma = 10^{-2}$ 
and different values of $n$ are shown for $\ell < 300$ 
(left panel) and for $\ell > 200$ (right panel).}
\label{fig:clT_delta_2}
\end{figure}

The power spectra of the CMB temperature anisotropies for different values of $\gamma$ are 
shown in the left panel of figure~\ref{fig:clT_delta}. The relative differences for CMB temperature 
anisotropies with respect to a $\Lambda$CDM reference 
model are shown in the right panel of figure~\ref{fig:clT_delta} for $\gamma=10^{-3}$, 
$\gamma=10^{-4}$ and in figure~\ref{fig:clT_delta_2} for different $n$ and $\gamma=10^{-2}$.

\begin{figure}[t!!]
\centering
\epsfig{file=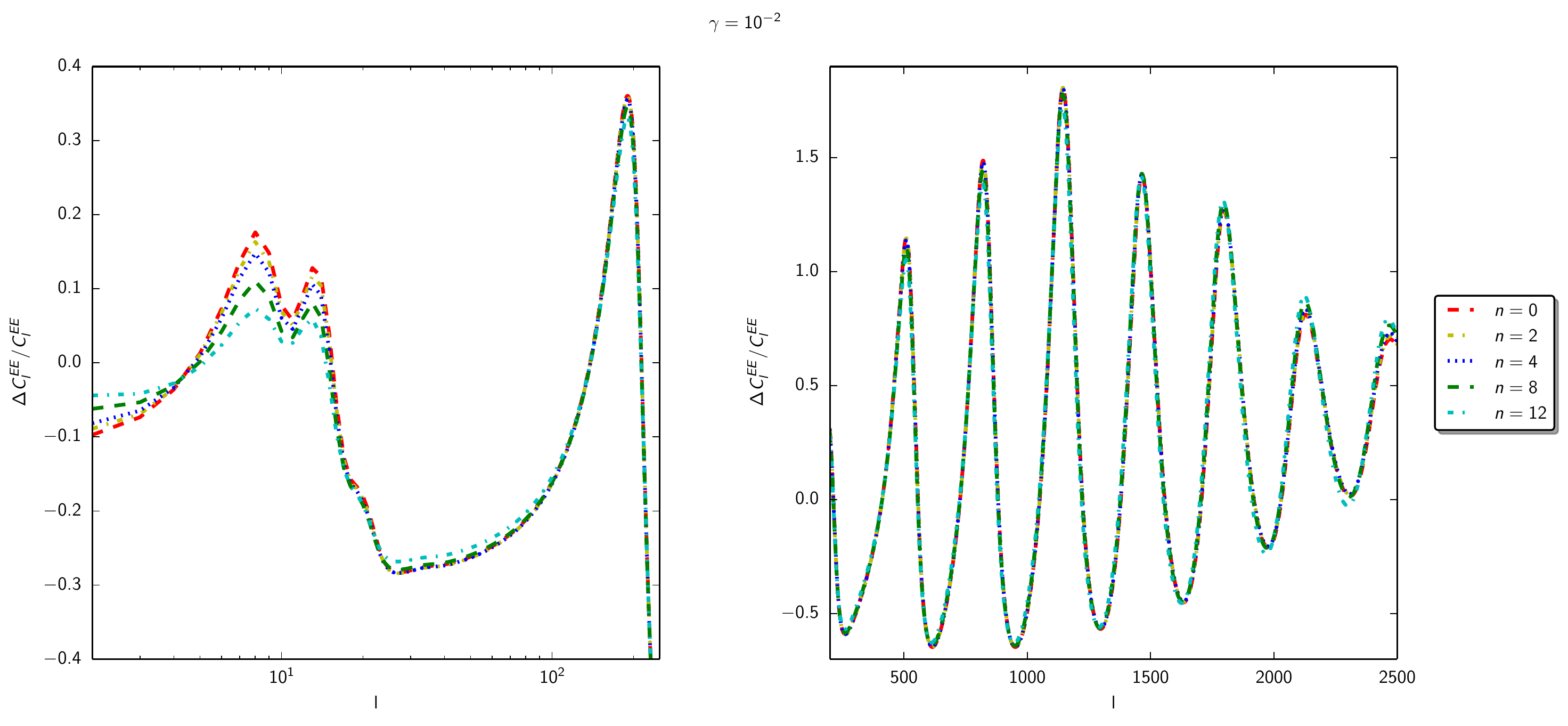, width=15 cm} \\
\caption{Relative differences for the CMB polarization E-mode anisotropies power spectrum  
with respect to a reference $\Lambda$CDM for $\gamma = 10^{-2}$ 
and different values of $n$ are shown for $\ell < 300$ 
(left panel) and for $\ell > 200$ (right panel).}
\label{fig:clE_delta}
\end{figure}

\begin{figure}[t!!]
\centering
\epsfig{file=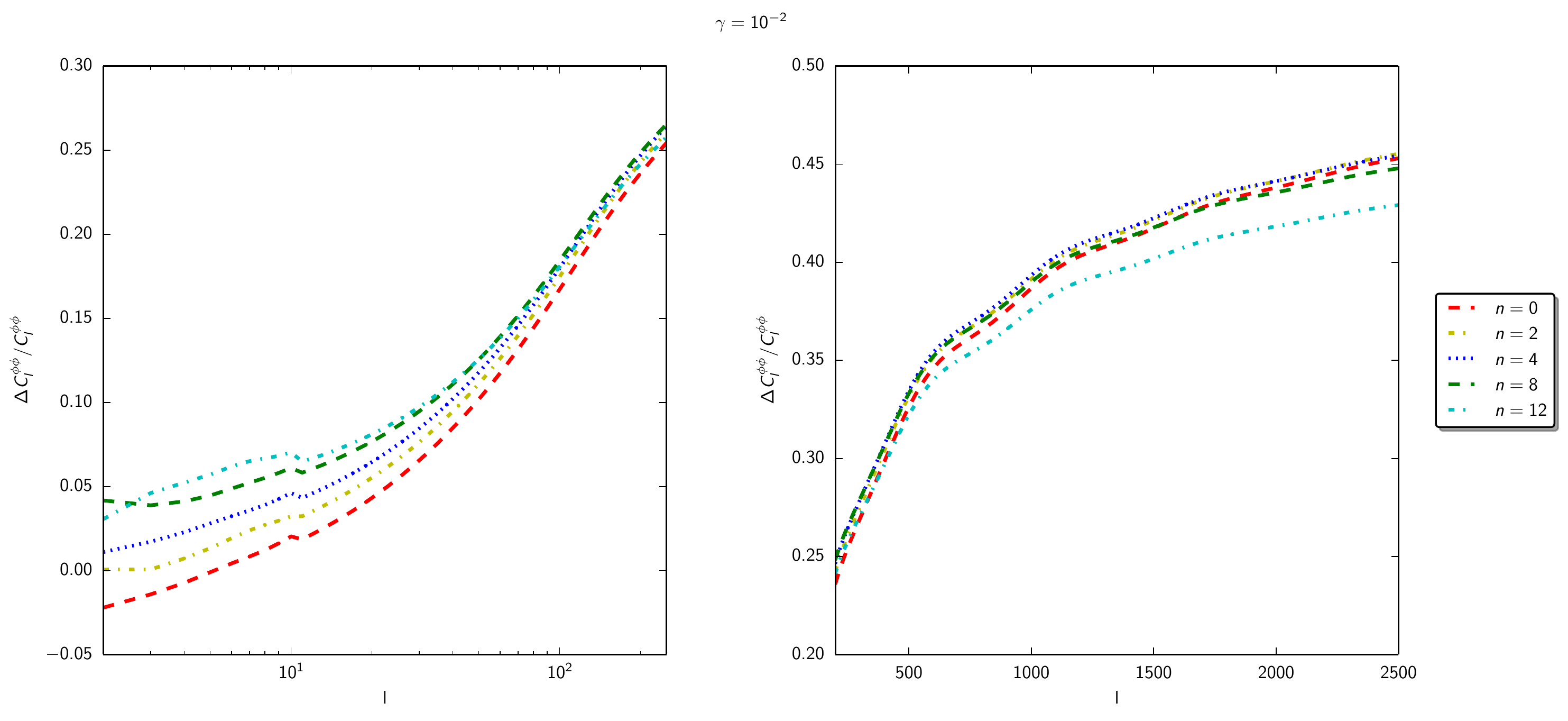, width=15 cm} \\
\caption{Relative differences for the CMB lensing power spectrum  
with respect to a reference $\Lambda$CDM for $\gamma = 10^{-2}$ 
and different values of $n$ are shown for $\ell < 300$ 
(left panel) and for $\ell > 200$ (right panel).}
\label{fig:clp_delta}
\end{figure}

\begin{figure}[t!!]
\centering
\epsfig{file=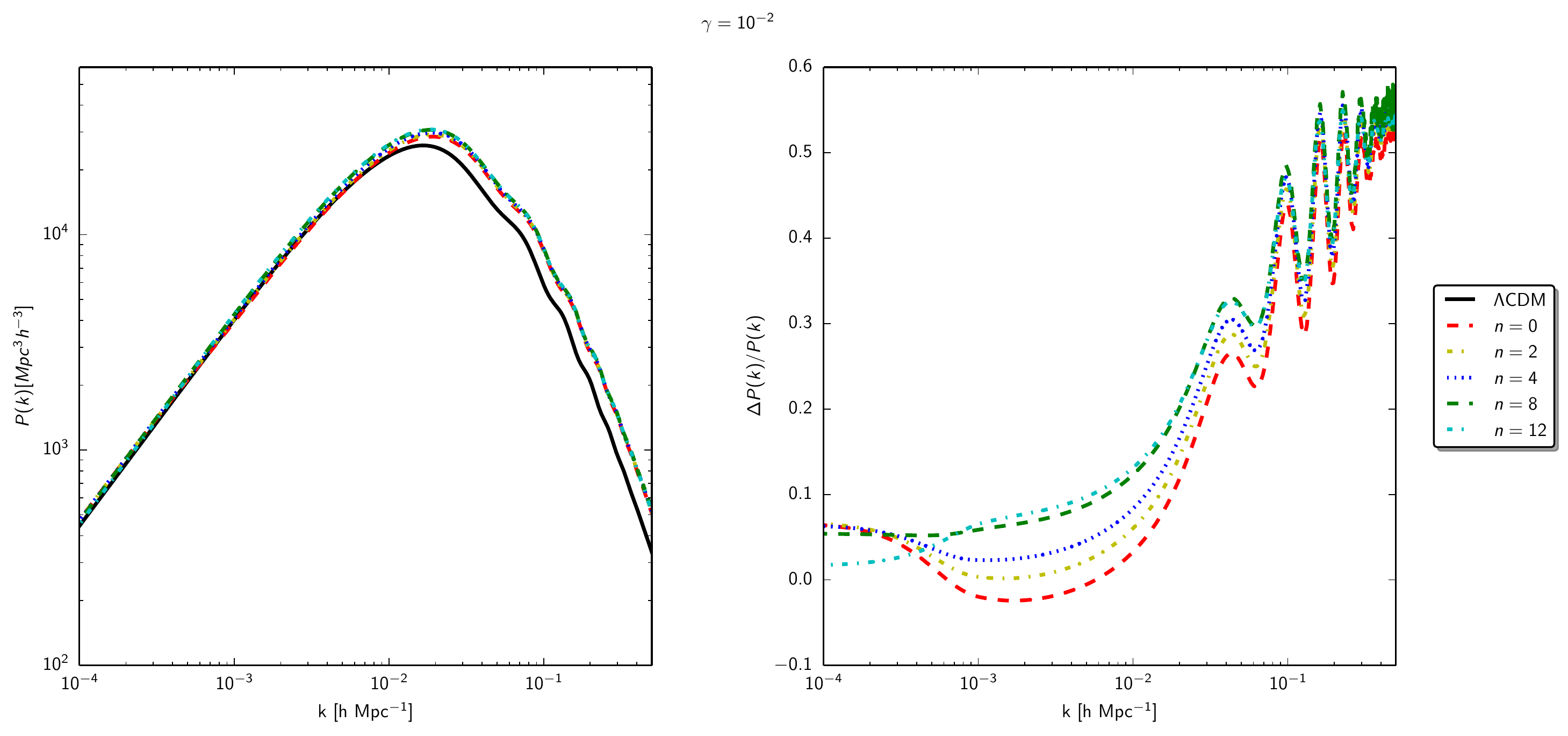, width=15 cm} \\
\caption{In the left panel, we show the linear matter power spectrum 
for $\gamma = 10^{-2}$ and different values of $n$. 
In the right panel, we show the relative difference for the linear matter power 
spectrum with respect to a reference $\Lambda$CDM.}
\label{fig:Pdk_delta}
\end{figure}

Analogous plots for the E-mode polarization, lensing and linear matter power spectrum 
(at $z=0$) are shown in figures~\ref{fig:clE_delta}, \ref{fig:clp_delta}, \ref{fig:Pdk_delta} 
for different $n$ and $\gamma=10^{-2}$. From these plots is clear that the impact of different 
$n$ is mainly relegated to $\ell \lesssim 30$ in the autocorrelator spectra of CMB 
temperature and polarization. The impact of different potentials is not negligible at 
smaller scales.
Instead, is not the case for CMB lensing (figure~\ref{fig:clp_delta}) and the linear matter 
power spectrum (figure~\ref{fig:Pdk_delta}).

\section{Updated $Planck$ 2015 results for the quartic potential}
\label{sec:Planck2015}

{
\input section_four_2.tex

}

\section{Cosmological constraints for $n\ne4$}
\label{sec:Planck2015pot}

{
\input section_five_2.tex

}

\section{Conclusions}
\label{sec:concl}

We have studied a simple class of modified gravity models alternative to $\Lambda$CDM, based on IG - 
or a Brans-Dicke-like - with a monomial potential. We have limited ourselves to positive values 
of the exponent, extending the case of a quartic potential previously studied in \cite{Umilta:2015cta}. 
In this class of models the scalar field increases from the constant value during the radiation 
era and therefore the effective Planck mass therefore increase in time during the matter 
dominated era.

Despite its semplicity, this class of models leads to distinct effects compared to $\Lambda$CDM for 
values of the coupling $\gamma$ compatible with observations, such as a slightly larger value of 
$H_0$ because of the modification of the expansion history due to the coupling of the scalar field 
to the Ricci curvature $\gamma$.
We have shown that the latter effect causes a shift of the CMB peaks similar to an open Universe. 
The dependence on the potential which drives the Universe in acceleration instead reveals itself 
in the Integrated Sachs-Wolfe effect, i.e. at $\ell \lesssim 20$, for CMB and in general at low 
redshifts for other cosmological probes, although to a much smaller extent. We have shown that 
the posterior probabilities for the standard cosmological parameters obtained with the most 
recent cosmological data are unchanged for different powers of the monomial potential. 
On the opposite, current values of $\dot G/G$ and $\ddot G/G$ (on cosmological scales) depend on 
the type of the potential: this dependence must be taken in mind when comparing the cosmological 
constraints with the Solar System constraints.

The full information of {\sc Planck} {\em alone}, i.e. temperature, polarization and lensing, is 
now capable to constrain $\gamma < 0.0017$ at 95~\% CL for $n=4$; by adding a compilation of BAO 
data this 95~\% CL constrant is further tightened to $\gamma < 0.00075$. With the increasing 
precision of cosmological observations the cosmological bounds on the variation of $G$ are 
fully consistent and closer to the Solar System constraints \cite{Bertotti:2003rm}.

Future accurate measurments from the $Euclid$ ESA mission \cite{Laureijs:2011gra,Amendola:2012ys} 
will further constrain $\gamma$ and will provide insights on the form of the potential 
in these induced gravity - or Brans-Dicke-like - dark energy models.

%Future accurate measurements of the history of the expansion of the Universe 
%from the $Euclid$ ESA mission \cite{Laureijs:2011gra,Amendola:2012ys} will help in studing 
%the potential in these class of induced gravity dark energy models which affects 
%the time evolution of the scalar field at low redshift.

\section*{Acknowledgements}
We wish to thank Julien Lesgourgues and Thomas Tram for many useful suggestions on the use 
of the CLASS code and Benjamin Audren for the useful suggestions on the use of the MONTEPYTHON code.
We wish to thank Jan Hamann for useful discussions on Big Bang nucleosynthesis.
The support by the "ASI/INAF 
Agreement 2014-024-R.0 for the Planck LFI Activity of Phase E2" is acknowledged.
We acknowledge financial contribution from the agreement ASI/INAF n. 
I/023/12/1.
MB acknowledges the TTK at RWTH University of Aachen for hospitality during writing part of 
this paper. CU has been supported within the Labex ILP (reference
ANR-10-LABX-63) part of the Idex SUPER, and received financial state
aid managed by the Agence Nationale de la Recherche, as part of the
programme Investissements d'avenir under the reference
ANR-11-IDEX-0004-02.

\end{document}

%% file: section_four_2.tex
In this and the following section we constrain this simple class of dark energy models with 
the {\sc Planck} \cite{Adam:2015rua,Plancklike15,Plancklensing15} and a compilation 
of BAO \cite{Beutler:2011hx,Ross:2014qpa,Anderson:2013zyy} data. 

We performed the Markov Chain Monte Carlo (MCMC) analysis by using the publicly available 
code {\sc Monte Python} 
\footnote{\href{https://github.com/baudren/montepython\_public}{https://github.com/baudren/montepython\_public}} \cite{Audren:2012wb} 
connected to our modified version of the code {\sc CLASS}.
We varied the six cosmological parameters for a flat $\Lambda$CDM model, i.e. 
$\omega_b$, $\omega_c$, $H_0$, $\tau$, $\ln\left(10^{10}\ A_s\right)$, $n_s$, plus one extra 
parameter related to the coupling with the Ricci curvature in Eq.~\eqref{eqn:IGaction}. 
We sampled on the quantity $\zeta$, according to \cite{Li:2013nwa, Umilta:2015cta}, defined as:
\begin{equation}
\zeta \equiv \ln \left( 1 + 4 \gamma \right) = \ln \left( 1 + \frac{1}{\omega_\mathrm{BD}}  \right)
\end{equation}
with the prior $[0,0.039]$. In the analysis we assumed 3 
massless neutrino and marginalized over foreground and calibration nuisance parameters which 
are also varied together with the cosmology.

As CMB data we use the {\sc Planck} 2015 release \cite{Plancklike15} based on the two-point function
as provided by 1. an exact pixel based likelihood at low resolution, which covers temperature and 
polarization data from $\ell=2$ to $29$ (the polarization part of this likelihood is denoted as 
"lowP" in the following), 2. a high-$\ell$ likelihood based on a Gaussian approximation available 
as temperature only or temperature plus polarization. We will refer in the following to $Planck$ 
TT (TT,TE,EE) as the combination of the TT (TT,TE,EE) likelihood at multipoles $\ell \ge 30$ and 
the low-$\ell$ temperature-only likelihood. 
We also use the $Planck$ 2015 lensing likelihood \cite{Plancklensing15}, in particular the 
version obtained from temperature and polarization data, with the multipole 
range $40\leq \ell\leq 400$.

In combination with CMB data we use measurments of $D_V/r_\textup{drag}$ 
by 6dFGRS at $z_\textup{eff}=0.106$ \citep{Beutler:2011hx}, SDSS-MGS at $z_\textup{eff}=0.15$ 
\citep{Ross:2014qpa}, SDSS-DR11 CMASS and LOWZ at $z_\textup{eff}=0.57$ and 
$z_\textup{eff}=0.32$ respectively \citep{Anderson:2013zyy}. Moreover, in section~\ref{sec:H0} 
we consider the impact of the local measurements on the Hubble's constant, i.e. 
$H_0 = 73.8 \pm 2.4$ km s$^{-1}$ Mpc$^{-1}$ \cite{Riess:2011yx} and 
$H_0 = 70.6 \pm 3.0$ km s$^{-1}$ Mpc$^{-1}$ \cite{Efstathiou:2013via}.

\begin{figure}[t!!]
\centering
\epsfig{file=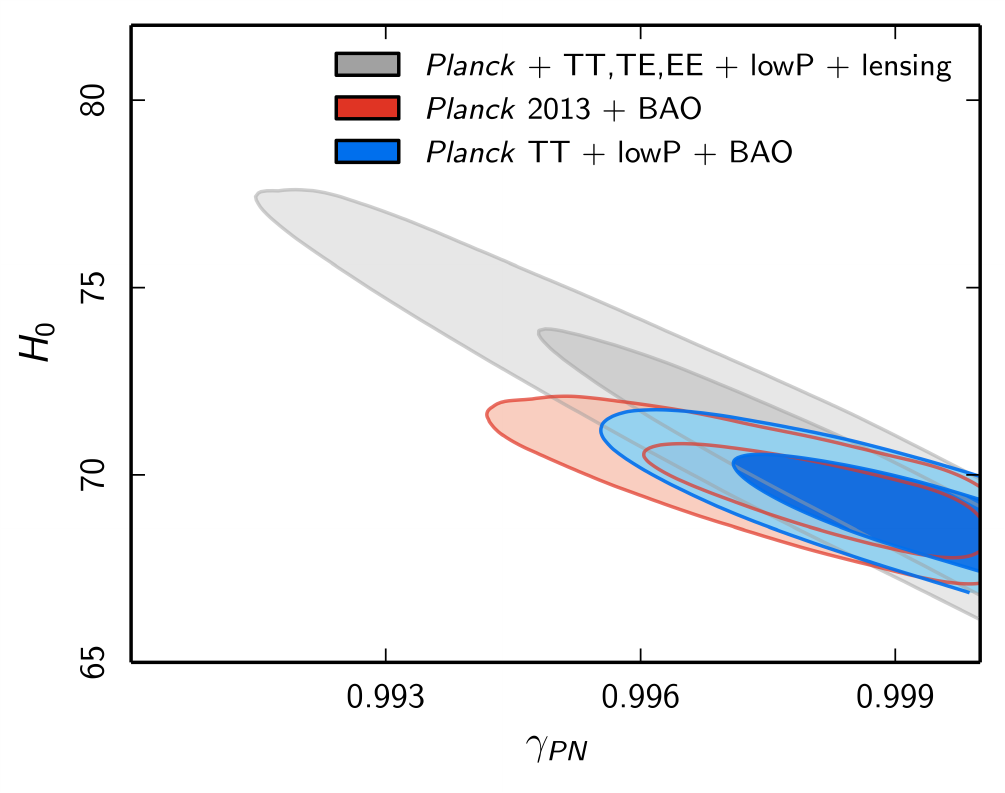, width=7.2 cm}
\epsfig{file=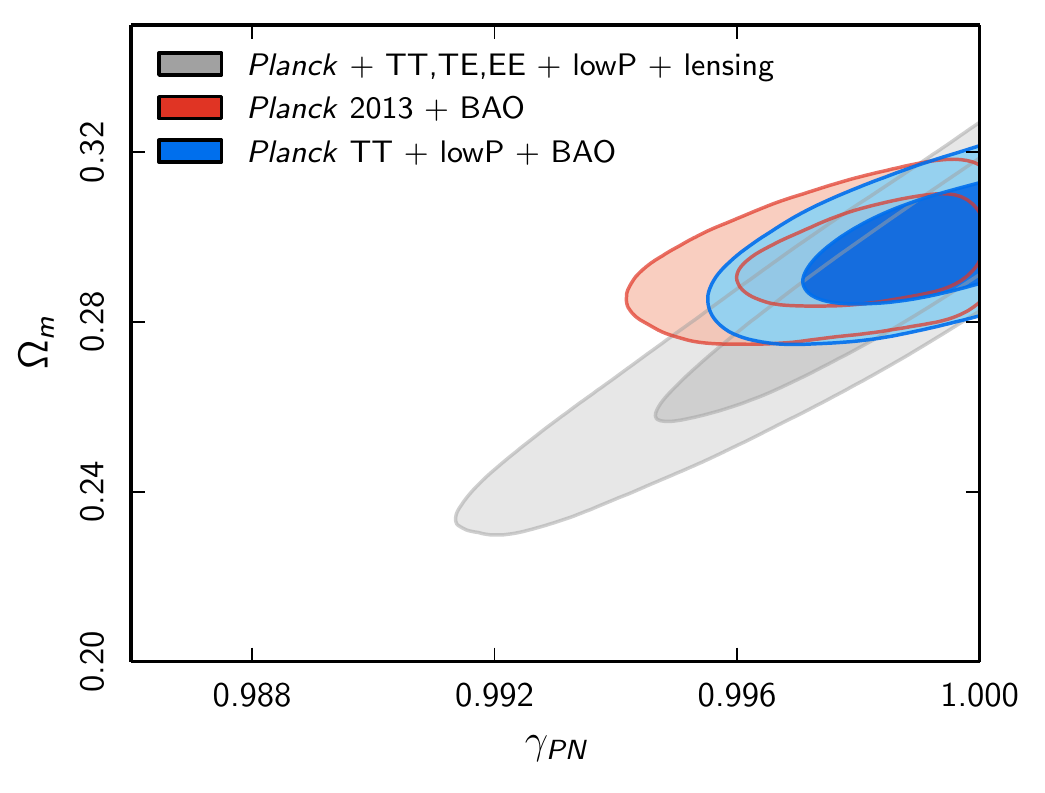, width=7.5 cm}
\caption{2-dimentional marginalized confidence levels at $68\,$\% and $95\,$\% for 
($\gamma_\textup{PN},\,H_0$) on the left and ($\gamma_\textup{PN},\,\Omega_\textup{m}$) on the 
right for $Planck$ TT,TE,EE + lowP + lensing (gray), $Planck$ 2013 + BAO (red) and 
$Planck$ TT + lowP + BAO (blue).}
\label{fig:2D}
\end{figure}

The constraints obtained from CMB and BAO data with $n=4$, for different combinations of data sets 
are summarized in table~\ref{tab:PlanckTT}.
These results update those presented in Ref.~\cite{Umilta:2015cta} based on the $Planck$ nominal 
mission temperature data, and use the same compilation of BAO data. See also 
\cite{Avilez:2013dxa,Li:2013nwa,Li:2015aug} for other works studying $Planck$ 2013 constraints on 
Brans-Dicke-like models. In combination the same BAO data, the full mission temperature data 
improve the $95\,$\% CL constraint on the coupling to the Ricci curvature $\gamma$ by $25\,$\% 
compared to the nominal mission data (see also figure~\ref{fig:2D}) to:
\be
\gamma < 0.00089 \, (95\,\%\ \text{CL, $Planck$ TT + lowP + BAO}) \,.
\label{gamma_basic}
\ee

\begin{figure}[t!!]
\centering
\epsfig{file=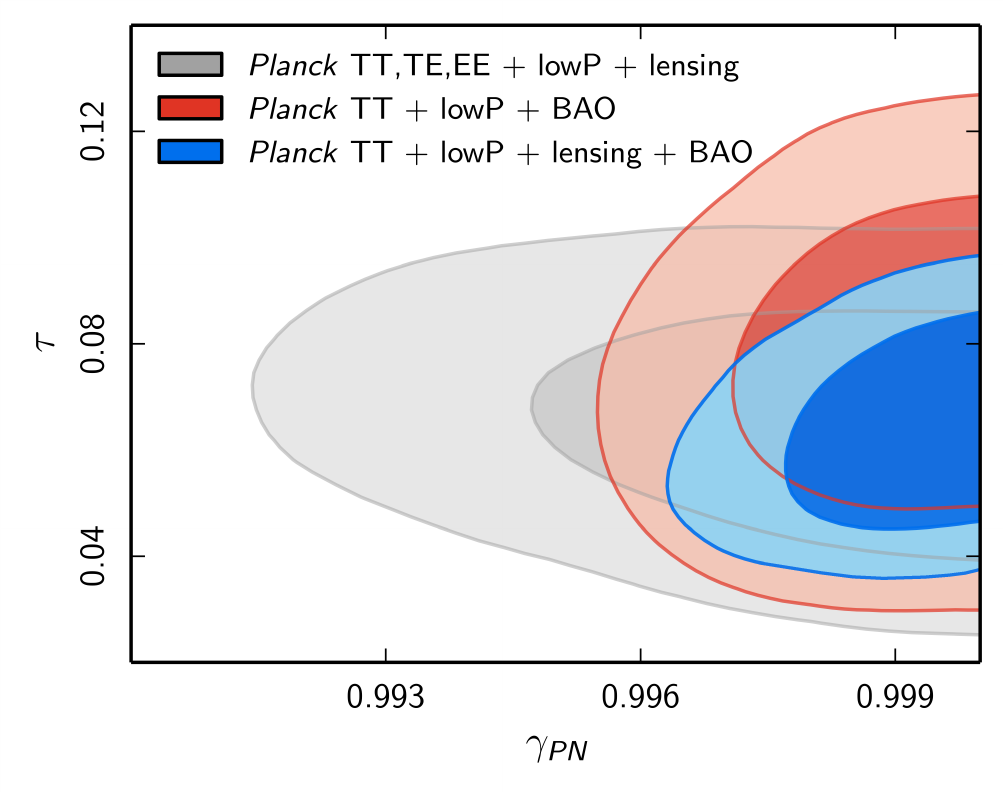, width=7.5 cm}
\epsfig{file=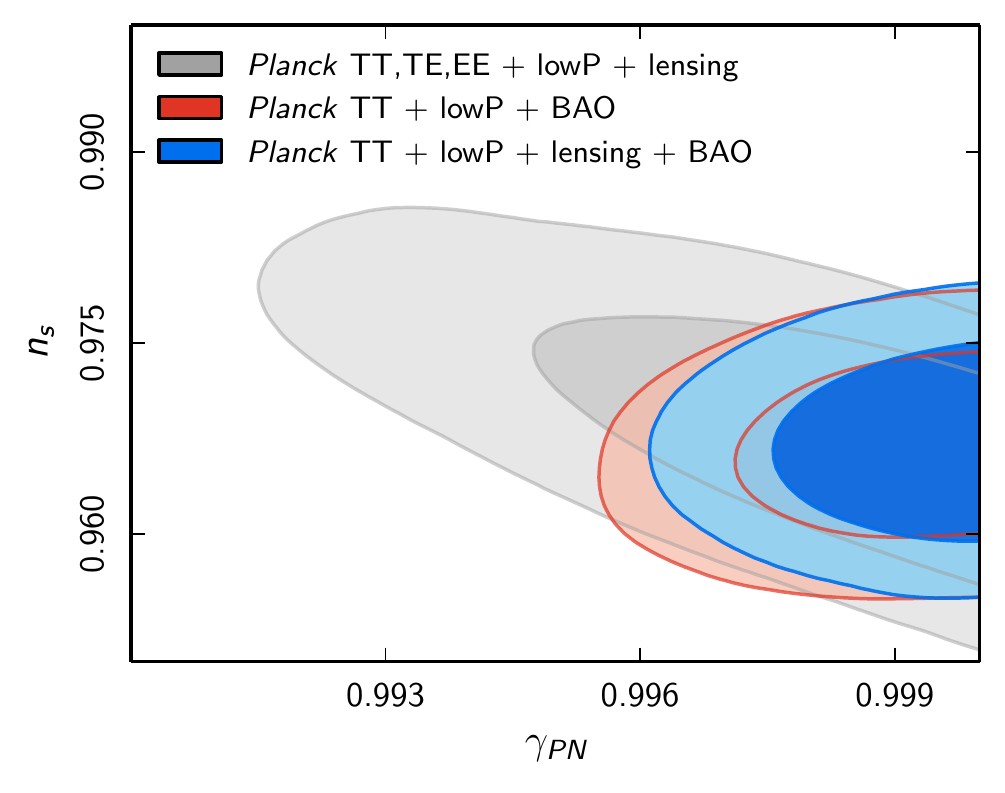, width=7.5 cm}
\caption{2-dimentional marginalized confidence levels at $68\,$\% and $95\,$\% for ($\gamma_\textup{PN},\,\tau$)
on the left and ($\gamma_\textup{PN},\,n_\textup{s}$) on the right for
$Planck$ TT,TE,EE + lowP + lensing (gray), $Planck$ 2013 + BAO (red) and
$Planck$ TT + lowP + BAO (blue).}
\label{fig:tau}
\end{figure}

We now discuss the impact of the $Planck$ lensing data \cite{Plancklensing15}.
One of the effects of the CMB lensing is to slightly favour smaller values of the amplitude of 
fluctuations $A_\mathrm{s}$ and therefore of the optical depth thanks to the accurate determination 
of $A_\mathrm{s} e^{-2 \tau}$ by the CMB temperature power spectrum measured by {\sc Planck}.
We show in figure~\ref{fig:tau} how the addition of $Planck$ lensing improves either the 
determination of $\tau$ and the constraint on $\gamma$.

\input table_params_T.tex

In table~\ref{tab:PlanckTTTEEE} we show the results with the inclusion of $Planck$ high-$\ell$ 
polarization data. The IG dark energy model with a quartic potential provide a better fit of the 
data compared to $\Lambda$CDM, but not at a statistically significant level - 
$\Delta \chi^2 \simeq -1.2$ for $Planck$ TT + lowP + BAO and $\Delta \chi^2 \simeq -2.3$ for 
$Planck$ TT,TE,EE + lowP + lensing. 
It is important to note that the full information of {\sc Planck} {\em alone}, i.e. temperature, 
polarization and lensing, is now capable to constrain $\gamma$:
\be
\gamma < 0.0017 \, (95\,\%\ \text{CL, $Planck$ TT,TE,EE + lowP + lensing}) \,.
\label{gamma_Planckonly_bound}
\ee
We quote the following $Planck$ TT,TE,EE + lowP + lensing + BAO at the 95\,\% CL constraint on the coupling 
to the Ricci curvature:
\be
\gamma < 0.00075 \, (95\,\%\ \text{CL, $Planck$ TT,TE,EE + lowP + lensing + BAO}) \,.
\label{gamma_tightbound}
\ee 

\input table_params_TP.tex

We quote also the derived constraints on the change of the effective Newton's constant between the 
radiation era and the present time $\delta G_\textup{N}/G_\textup{N} \equiv (\sigma_i^2 - \sigma^2_0)/\sigma_0^2$:
\be
\frac{\delta G_\textup{N}}{G_\textup{N}} = -0.002_{-0.037}^{+0.002}\ (95\,\%\ \text{CL, $Planck$ TT,TE,EE + lowP + BAO})
\label{deltaG_bound}
\ee
and the constraint on its derivatives ($\dot G_\textup{N}/G_\textup{N} \equiv - 2 \dot \sigma_0 / \sigma_0$) 
at present time:
\begin{align}
&\frac{\dot{G}_\textup{N}}{G_\textup{N}} (z=0) = -0.08_{-0.55}^{+0.08}\ [\times10^{-13}\ \text{yr}^{-1}]\, (95\,\%\ \text{CL, $Planck$ TT,TE,EE + lowP + BAO}) \,,
\label{eqn:dotGbound}\\
&\frac{\ddot{G}_\textup{N}}{G_\textup{N}} (z=0) = 0.36_{-0.36}^{+0.26}\ [\times10^{-23}\ \text{yr}^{-1}]\, (95\,\%\ \text{CL, $Planck$ TT,TE,EE + lowP + BAO}) \,.
\label{eqn:ddotGbound}
\end{align}

The constraints derived here are the tighter obtained from cosmological observations for similar 
scalar-tensor models (for a comparison, the $95\,$\% CL constraint from pulsar timing is 
$\dot{G}_\textup{N}/G_\textup{N} = -0.6\pm 1.1 \times 10^{-12}\ \text{yr}^{-1}$ \cite{Zhu:2015mdo}).

\subsection{Combination with local measurements}
\label{sec:H0}

As in \cite{Umilta:2015cta} we analyze the combination of the local measurements of the Hubble constant with 
$Planck$ TT + lowP by considering the impact of two different local estimates of $H_0$, 
such as:
$H_0 = 73.8 \pm 2.4$ km s$^{-1}$ Mpc$^{-1}$ \cite{Riess:2011yx}, denoted as $H_0^*$, and 
$H_0 = 70.6 \pm 3.0$ km s$^{-1}$ Mpc$^{-1}$ \cite{Efstathiou:2013via}, denoted as $H_0^\dagger$.
We find:
\begin{align}
&H_0 = 73.1^{+2.1}_{-2.3}\ [\text{km}\ \text{s}^{-1}\ \text{Mpc}^{-1}]\ (68\,\%\ \text{CL, $Planck$ TT + lowP + H$_0^*$}) \\
&\gamma = 0.0011\pm{0.0010}\ (95\,\%\ \text{CL, $Planck$ TT + lowP + H$_0^*$})\,,
\end{align}
and
\begin{align}
&H_0 = 71.3^{+1.8}_{-2.8}\ [\text{km}\ \text{s}^{-1}\ \text{Mpc}^{-1}]\ (68\,\%\ \text{CL, $Planck$ TT + lowP + H$_0^\dagger$}) \\
&\gamma < 0.0017\ (95\,\%\ \text{CL, $Planck$ TT + lowP + H$_0^\dagger$})\,.
\end{align}
We note that the degeneracy of $H_0$ with higher value of $\gamma$ has been reduced 
with the improved accuracy of the $Planck$ full mission temperature data, compared 
to the nominal mission data \cite{Umilta:2015cta}.

In combination with BAO, we find:
\be
H_0 = 69.4^{+0.8}_{-1.0}\ [\text{km}\ \text{s}^{-1}\ \text{Mpc}^{-1}]\ (68\,\%\ \text{CL, $Planck$ TT + lowP + BAO}) \,,
\ee
which is larger than the value obtained for the $\Lambda$CDM model with three massless neutrinos,
i.e. $67.8 \pm 0.6$ km s$^{-1}$ Mpc$^{-1}$, for the same combination of datasets.

\subsection{BBN consistency relation on $G_\textup{N}$}

The value of the effective gravitational 
constant determines the expansion rate in the radiation era and therefore can affect 
the cosmological abundances of the light elements during Big Bang Nucleosynthesis (BBN). Therefore, 
BBN was used to provide limits to the variation 
of the effective Newton's constant \cite{Copi:2003xd, Bambi:2005fi}.

In the following we investigate the impact of 
the modification of the BBN consistency condition implemented in the public 
code {\sc PArthENoPE} \cite{Pisanti:2007hk} due to the different value of the effective Newton's constant during nucleosynthesis.
We consider the effect of a different gravitational constant 
as a source of extra radiation in $Y_\textup{P}^\textup{BBN}(\omega_b,\,N_\textup{eff})$, 
\cite{Hamann:2007sb}. 
It is interesting to note that with this improved BBN consistency condition the posterior 
probabilities for the primary cosmological parameters are unaffected, and we just observe a 
small shift for the primordial Helium abundance towards higher values.

%% file: table_params_T.tex
\begin{table*}
\centering{\scriptsize
\begin{tabular}{|l|ccc|}
\hline
                                              & $Planck$~2013               & $Planck$~TT + lowP           & $Planck$~TT + lowP                 \\
                                              & + BAO                       & + BAO                        & + lensing + BAO                 \\
\hline
$ 10^5\Omega_\mathrm{b}h^2$                   & $2203\pm25$                 & $2224\pm21$                  & $2224_{-21}^{+20}$            \\
$ 10^4\Omega_\mathrm{c}h^2$                   & $1207_{-22}^{+18}$          & $1198_{-17}^{+16}$           & $1191\pm14$                     \\
$H_0$ [km s$^{-1}$ Mpc$^{-1}$]                & $69.5_{-1.2}^{+0.9}$        & $69.4_{-1.0}^{+0.8}$         & $69.4_{-0.9}^{+0.7}$           \\
$\tau$                                        & $0.088_{-0.013}^{+0.012}$   & $0.076_{-0.018}^{+0.019}$    & $0.063_{-0.014}^{+0.012}$      \\
$\ln \left(  10^{10} A_\mathrm{s} \right)$    & $3.090_{-0.026}^{+0.024}$   & $3.087\pm{0.036}$            & $3.059_{-0.026}^{+0.022}$         \\
$n_{\mathrm s}$                               & $0.9611 \pm 0.0053$         & $0.9665\pm0.0046$            & $0.9669_{-0.0047}^{-0.0042}$     \\
$\zeta$                                       & $<0.0047$ (95\% CL)         & $<0.0036$ (95\% CL)          & $<0.0031$ (95\% CL)        \\
\hline
$10^3 \gamma$                                 & $<1.2$ (95\% CL)            & $<0.89$ (95\% CL)            & $<0.75$ (95\% CL)     \\
$\gamma_{PN}$                                 & $>0.9953$ (95\% CL)         & $>0.9965$ (95\% CL)          & $>0.9970$ (95\% CL)     \\
$\Omega_\mathrm{m}$                           & $0.295\pm0.009$             & $0.295\pm0.008$              & $0.294\pm0.008$        \\
$\delta G_\mathrm{N}/G_\mathrm{N}$            & $-0.015_{-0.006}^{+0.013}$  & $-0.011_{-0.004}^{+0.010}$   & $-0.009_{-0.009}^{+0.003}$     \\
$10^{13} \dot{G}_\mathrm{N}(z=0)/G_\mathrm{N}$ [yr$^{-1}$]   & $-0.61_{-0.25}^{+0.55}$  & $-0.45_{-0.16}^{+0.43}$  & $-0.37_{-0.12}^{+0.34}$   \\
$10^{23} \ddot{G}_\mathrm{N}(z=0)/G_\mathrm{N}$ [yr$^{-2}$]  & $0.86_{-0.78}^{+0.33}$   & $0.63_{-0.58}^{+0.22}$   & $0.52_{-0.50}^{+0.17}$    \\
\hline
\end{tabular}}
\caption{Constraints on main and derived parameters for $Planck$ TT + lowP + BAO (at 68\% CL if not otherwise stated). 
In the first column we report the results obtained with the previous $Planck$ 2013 data from Ref.~\cite{Umilta:2015cta}}
\label{tab:PlanckTT}
\end{table*}

%% file: table_params_TP.tex
\begin{table*}
\centering{\scriptsize
\begin{tabular}{|l|ccc|}
\hline
                                              & $Planck$ TT,TE,EE             & $Planck$ TT,TE,EE             & $Planck$ TT,TE,EE                  \\
                                              & + lowP + lensing              & + lowP + BAO                  & + lowP + lensing + BAO      \\
\hline
$ 10^5\Omega_\mathrm{b}h^2$                   & $2234\pm17$                 & $2231\pm14$                  & $2223\pm20$               \\
$ 10^4\Omega_\mathrm{c}h^2$                   & $1189\pm14$                 & $1194\pm12$                  & $1191_{-14}^{+15}$                      \\
$H_0$ [km s$^{-1}$ Mpc$^{-1}$]                & $71.0_{-3.0}^{+1.4}$        & $69.4_{-1.1}^{+0.6}$         & $69.4_{-1.0}^{+0.5}$  \\
$\tau$                                        & $0.066_{-0.013}^{+0.012}$   & $0.079_{-0.016}^{+0.017}$    & $0.063_{-0.014}^{+0.012}$ \\
$\ln \left(  10^{10} A_\mathrm{s} \right)$    & $3.066_{-0.028}^{+0.024}$   & $3.095_{-0.033}^{+0.031}$    & $3.059_{-0.026}^{+0.021}$ \\
$n_{\mathrm s}$                               & $0.9695 \pm 0.0056$         & $0.9675\pm0.0041$            & $0.9669_{-0.0048}^{-0.0043}$ \\
$\zeta$                                       & $<0.0068$ (95\% CL)         & $<0.0030$ (95\% CL)          & $<0.0030$ (95\% CL)    \\
\hline
$10^3 \gamma$                                 & $<1.7$ (95\% CL)            & $<0.76$ (95\% CL)            & $<0.75$ (95\% CL)        \\ 
$\gamma_{PN}$                                 & $>0.9933$ (95\% CL)         & $>0.9970$ (95\% CL)          & $>9970$ (95\% CL)          \\
$\Omega_\mathrm{m}$                           & $0.281\pm0.009$             & $0.295\pm0.015$              & $0.294\pm0.008$              \\ 
$\delta G_\mathrm{N}/G_\mathrm{N}$            & $-0.020_{-0.005}^{+0.019}$  & $-0.010_{-0.009}^{+0.004}$   & $-0.009_{-0.009}^{+0.003}$    \\
$10^{13} \dot{G}_\mathrm{N}(z=0)/G_\mathrm{N}$ [yr$^{-1}$]   & $-0.77_{-0.27}^{+0.43}$  & $-0.39_{-0.15}^{+0.35}$  & $-0.37_{-0.12}^{+0.34}$   \\
$10^{23} \ddot{G}_\mathrm{N}(z=0)/G_\mathrm{N}$ [yr$^{-2}$]  & $1.2_{-1.1}^{+0.4}$   & $0.56_{-0.50}^{+0.21}$   & $0.52_{-0.50}^{+0.17}$   \\
\hline
\end{tabular}}
\caption{\label{tab:PlanckTTTEEE} Constraints on main and derived parameters for $Planck$ TT,TE,EE + lowP
with different combination of other datasets (at 68\% CL if not otherwise stated).}
\end{table*}

%% file: section_five_2.tex
\begin{figure}[t!!]
\centering
\epsfig{file=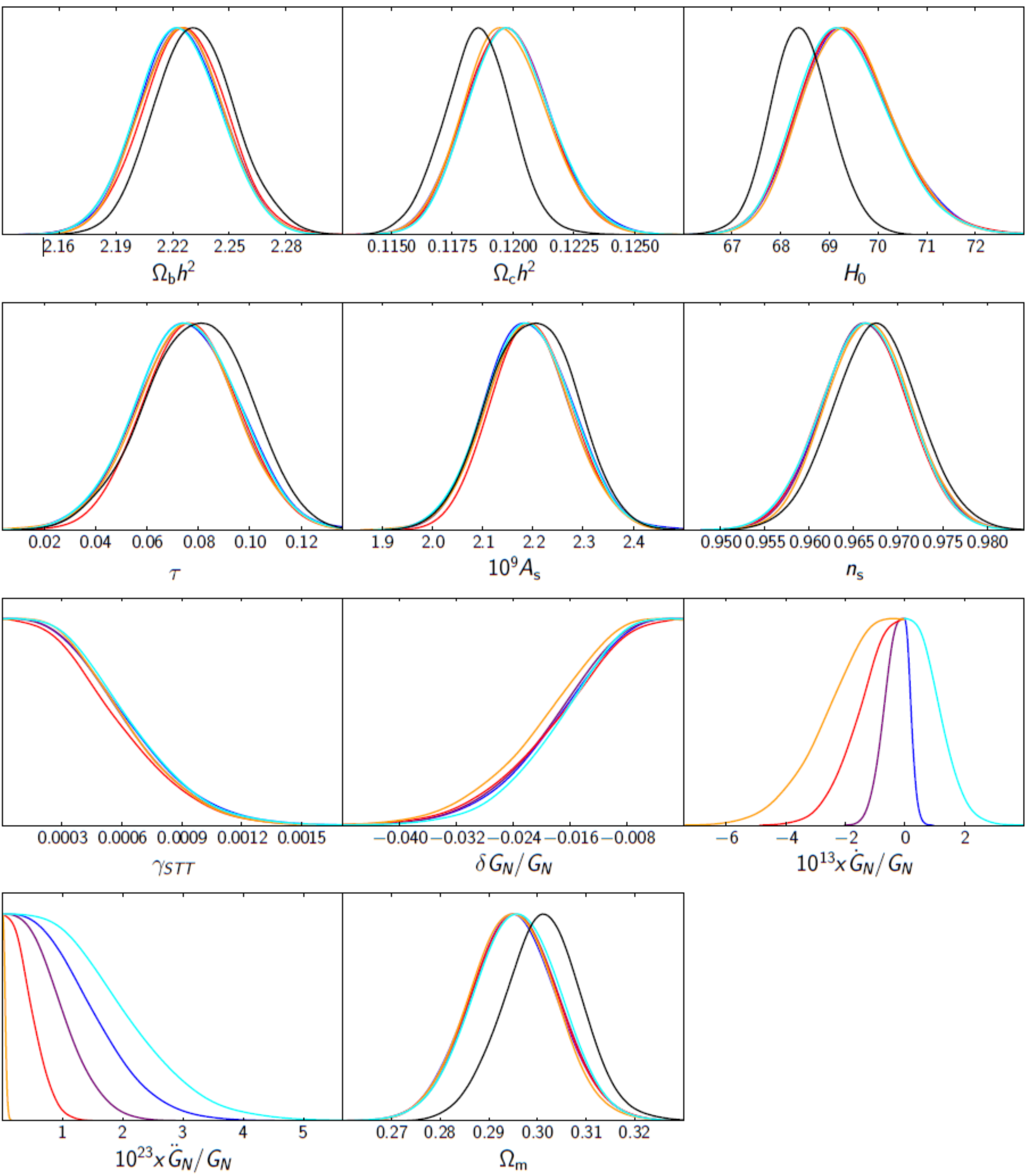,width=16cm}
\caption{1-dimentional likelihoods for the cosmological parameters for IG with different index of the 
monomial potential with $Planck$ TT + BAO, i.e. n = 0 (orange), 2 (red), 4 (purple), 6 (blue) and 8 (cyan). We show the 
comparison w.r.t. the $\Lambda$CDM cosmologigy plotted in black.}
\label{fig:Gpot}
\end{figure}

In this section we explore the impact of different $n$ on the cosmological parameters. 
Current data cannot discriminate at a statistical significant level between different values of $n$ and 
we therefore fix $n$ to representative values ($n=0,2,6,8$) 
and we vary the seven primary cosmological parameters (as well as the other foreground/nuisance 
parameters) as done in the previous section. Figure~\ref{fig:Gpot} shows that the posterior probabilities for the seven primary cosmological parameters 
hardly change for these different values of $n$: these results are compatible with the dependence of CMB anisotropies 
on $n$ shown in figures~\ref{fig:clT_delta}, \ref{fig:clE_delta}. Therefore, this class of scalar-tensor models is compatible with current 
cosmological data for a larger value of $H_0$ and a slightly smaller value for $\Omega_\mathrm{m}$. 

On the opposite, we note that the posterior probabilities for 
$\dot G_\textup{N}/G_\textup{N}$ and $\ddot G_\textup{N}/G_\textup{N}$ at present time depend on $n$. 
The dependence of the bounds $\dot G_\textup{N}/G_\textup{N}$ on $n$ can be easily 
understood from Eq.~\ref{fig:clT_delta}: for $n = 6$ and $n = 8$, the derivative of the scalar field 
becomes negative at the onset of the 
accelerated stage.  This dependence on $n$ must be kept in mind when comparing cosmological bounds 
on $\dot G_\textup{N}/G_\textup{N} (z=0) $ and $\ddot G_\textup{N}/G_\textup{N} (z=0)$ 
(which depend on the form of the potential) with Solar System constraints on the same 
time variations (which are obtained extrapolating from the massless case since the effect of the potential is considered negligible 
on such smaller scales detached from the cosmological expansion).

The bound on the shift of the scalar field between today and the radiation era is the same however 
it's evolution show a strong dependence from the choice of the potential as summarized in table~\ref{tab:G_n}.
\input table_params_n.tex

%% file: table_params_n.tex
\begin{table*}
\centering{\scriptsize
\begin{tabular}{|l|ccccc|}
\hline
                                                             & $n=0$       &   $n=2$       &   $n=4$       &   $n=6$       &   $n=8$      \\
\hline
$\delta G_\mathrm{N}/G_\mathrm{N}$                           & $>-0.028$   &   $>-0.027$   &   $>-0.026$   &   $>-0.026$   &   $>-0.025$     \\
$10^{13} \dot{G}_\mathrm{N}(z=0)/G_\mathrm{N}$ [yr$^{-1}$]   & $>-3.9$     &   $>-2.5$     &   $>-0.11$    &   $<0.4$      &   $<1.9$   \\
$10^{23} \ddot{G}_\mathrm{N}(z=0)/G_\mathrm{N}$ [yr$^{-2}$]  & $<0.077$    &   $<0.78$     &   $<1.5$      &   $<2.3$      &   $<3.0$    \\
\hline
\end{tabular}}
\caption{Constraints on the variation of the gravitational constant and its time derivatives (at 95\% CL) 
for different values of $n$ with $Planck$ TT + lowP + BAO}\label{tab:G_n}
\end{table*}